\documentclass{aa}  
\usepackage{graphicx}
\usepackage{color}
\usepackage{gensymb}
\usepackage{textcomp}
\usepackage[flushleft]{threeparttable}
\usepackage{amssymb}
\usepackage{txfonts}
\begin{document}

   \title{Spectroscopy of the helium-rich binary ES Ceti reveals accretion via a disc and evidence for eclipses}


   \author{K. B\k{a}kowska
          \inst{1}
          \and
          T. R. Marsh\inst{2}
          \and 
          D. Steeghs\inst{2}
          \and
          G. Nelemans\inst{3,4,5}
          \and
          P. J. Groot\inst{3,6,7,8}
          }

\institute{Institute of Astronomy, Faculty of Physics, Astronomy and Informatics, Nicolaus Copernicus University, ul. Grudzi\k{a}dzka 5,\\ 87-100 Toru\'{n}, Poland\\
        \email{bakowska@umk.pl}
    \and
    Department of Physics, University of Warwick, Coventry CV4 7AL, United Kingdom
    \and
    Department of Astrophysics/IMAPP, Radboud University, P O Box 9010, NL-6500 GL Nijmegen, The Netherlands
    \and
    Institute of Astronomy, KU Leuven, Celestijnenlaan 200D, B-3001 Leuven, Belgium
    \and
    SRON, Netherlands Institute for Space Research, Sorbonnelaan 2, NL-3584 CA Utrecht, The Netherlands
    \and
    Department of Astronomy, University of Cape Town, Private Bag X3, Rondebosch, 7701, South Africa
    \and
    South African Astronomical Observatory, P.O. Box 9, Observatory, 7935, South Africa
    \and
    The Inter-University Institute for Data Intensive Astronomy, University of Cape Town, Private Bag X3, Rondebosch, 7701, South Africa
}

   \date{Received: XXX; accepted: YYY}

 
  \abstract
   {Amongst the hydrogen-deficient accreting binaries known as the "AM~CVn stars" are three systems with the shortest known orbital periods: HM~Cnc (321\,s), V407~Vul (569\,s) and ES~Cet (620\,s). These compact binaries are predicted to be strong sources of persistent gravitational wave radiation. HM~Cnc and V407~Vul are undergoing direct impact accretion in which matter transferred from their donor hits the accreting white dwarfs directly. ES~Cet, is the longest period of the three and is amongst the most luminous AM~CVn stars, but it is not known whether it accretes via a disk or direct impact. ES~Cet displays strong HeII~4686 line emission, which is sometimes a sign of magnetically-controlled accretion.
   Peculiarly, although around one third of hydrogen accreting white dwarfs show evidence for magnetism, none have been found amongst helium accretors.
   }
   {We present the results of Magellan and VLT spectroscopic and  spectropolarimetric observing campaigns dedicated to ES~Cet with the aim of understanding its accretion structure.
   }
   {Based on the data collected, we derived trailed spectra, computed Doppler maps of the emission lines, and looked for circular polarisation and variability.}
   {We find strong variability in our spectra on the 620\,s period. The lines show evidence for double-peaked emission, characteristic for an accretion disc, with an additional component associated with the outermost disc, rather than a direct impact, that is broadly consistent with "S"-wave emission from the gas stream/disc impact region. This confirms beyond any doubt that 620\,s is the orbital period of ES~Cet. We find no significant circular polarisation (below $0.1$\,\%). The trailed spectra show that ES~Cet's outer disc is eclipsed by the mass donor, revealing at the same time that the photometric minimum coincides with the hitherto unrecognised eclipse.}
   {ES~Cet shows spectroscopic behaviour consistent with accretion via a disc, and is the shortest orbital period eclipsing AM~CVn star known.}


   \maketitle
%
%

\section{Introduction}

AM~CVn stars are ultra-compact binary systems with orbital periods of 5-65 minutes in which white dwarfs accrete from degenerate or semi-degenerate companions (recent reviews in \citealt{Solheim2010, Ramsay2018}). Only hydrogen-deficient material can attain the density necessary to fit within Roche lobes at such short orbital periods and the spectra of AM~CVn stars are devoid of hydrogen lines, and are dominated instead by helium, sometimes along with with heavier elements(see \citealt{Warner1995} for a detailed review). 

AM~CVn stars and cataclysmic variables (CVs) share many common features, including accretion discs which undergo semi-regular cycles of outburst and quiescence in some cases, and features characteristic of these discs are seen in the spectra of both classes.
There are also differences between the two types; obviously their abundances and orbital periods are very different, but two other distinctions can be drawn. As yet there are no known examples of magnetic white dwarfs amongst almost 60 known AM~CVn stars whereas around one third of CVs are magnetic \citep{Pala2020}; there is no obvious explanation for this difference.  The other difference is that AM~CVn stars can reach such short periods that a novel form of accretion becomes possible in which no accretion disc forms but matter directly impacts the accreting white dwarf instead \citep{Marsh2002,Marsh2004}.

For decades following the recognition of the nature of AM~CVn itself \citep{Smak1967}, only a few members of the class were identified. Even by the year 2000, just six systems were known, but wide-field spectroscopic and photometric surveys have since had a significant impact. By the time of \cite{Solheim2010} 25 systems were known, while 57 were listed by \cite{Ramsay2018}. Because of their short orbital periods, the evolution of AM~CVn stars is expected to be governed by gravitational-wave radiation \citep{Paczynski1967}. The radiation emitted from this hydrogen-deficient class of objects should be detectable by instruments such as the Laser Interferometer Space Antenna (LISA), hence they are test-beds of gravitational wave physics \citep{Kupfer2018}. AM~CVn stars are  also of interest as a potential progenitor class of Type~Ia supernovae  \citep{Brown2011,Gilfanov2010,Shen2014} and
can offer important insights into binary evolution and common-envelope evolution.

Three channels of formation for AM~CVn stars have been proposed. One possibility is via a double degenerate system that initially starts out as detached and then evolves to become interacting via mass transfer. However, this path has been in doubt since \cite{Shen2015} found that it was vulnerable to mergers caused by classical-nova-like eruptions from the accreting white dwarf. Another channel is via a semi-degenerate helium (He) star channel, while a third is to evolve from a hydrogen-rich CV (see e.g. \citealt{Nelemans2001a, Nelemans2001b, Podsiadlowski2003, Yungelson2008, Brooks2015}). We currently face a conundrum in understanding the origin of helium-rich CVs because all identified pathways to their formation face problems when confronted with observations, for example the recent study of the eclipsing AM~CVn, Gaia~14aae, \citep{Green2019}.

ES Ceti (ES~Cet) was discovered by \cite{Noguchi1980} and classified as a CV by \cite{Downes1993}. The optical magnitude range of the star is $16.5-16.8$, and with an absolute magnitude of $M_g=5.6\pm0.4$, ES~Cet is amongst the most luminous AM~CVn stars \citep{Ramsay2018}. \cite{Warner2002} found a photometric period in optical photometry of ES~Cet of 620\,s. Photometric studies over a baseline of years showed the period to be highly coherent, albeit slowly increasing with time (\citealt{Espaillat2005,Copperwheat2011,deMiguel2018}). Combined with the helium-dominated emission line spectrum, it was natural to interpret the period as orbital. ES~Cet has the third shortest orbital period currently known amongst AM~CVn stars, after HM~Cnc ($P_{orb}=324$\,s, \citealt{Roelofs2010}) and V407~Vul ($P_{orb}=569$\,s, \citealt{Haberl1995}). ES~Cet's period increases  at a rate consistent with expectations for mass transfer driven by gravitational radiation \citep{deMiguel2018}. However, the orbital periods of both HM~Cnc and V407~Vul decrease (\citealt{Esposito2014} and references therein). This was shown to be possible if the donors in these stars still have significant surface hydrogen layers as a result of the prior evolution \citep{DAntona2006, Kaplan2012}. There is indeed some evidence that hydrogen contributes to the emission line spectrum of HM Cnc \citep{Roelofs2010}. ES~Cet may differ from these systems and is possibly more representative of the bulk of AM~CVn stars.

Depending upon the size of the accretor in a mass-exchanging binary, it is possible that a disc is unable to form and instead the mass transferred ploughs directly into the accretor. This is familiar for main-sequence stars in the form of Algol binary stars. The same only becomes possible for typical white dwarfs at periods below $\sim10$\,min \citep[e.g.][]{Marsh2002, Roelofs2010}. Such direct impact is likely to lead to strong asymmetries visible in both photometry and spectroscopy. On the other hand if a disc forms, it tends to be relatively axi-symmetric, albeit with a disturbance caused by the gas stream at the edge of the disc. In the case of discs we see broad absorption lines in some high-state systems or double-peaked line emission from the disc plus a feature associated with the gas stream/disc impact which executes a sinusoidal radial velocity curve with a full amplitude similar to the separation between the peaks  \citep[e.g.][]{Marsh1999, Morales2003}. HM~Cnc and V407~Vul both appear to be in a state of direct impact \citep{Barros2007}. The next known AM~CVn star, with a longer orbital period than ES~Cet, SDSS J135154.46-064309.0 ($P_{orb} = 15.7$ min, \citealt{Green2018}) is disc-accreting system. ES~Cet's period is close to the likely dividing line between direct impact and disc accretion and so its mode of accretion is not clear \emph{ab initio}. 
Based on an analysis of the light curves, \cite{Espaillat2005} suggested that in fact ES~Cet might be in a state of direct-impact accretion as well, but this issue has not been settled in the years since.

The issues outlined above motivated us to carry out a spectroscopic study of ES~Cet. Spectroscopy moreover is a means to confirm the basic hypothesis that the 620\,s photometric period is indeed the binary period. Likely though this seems, given the nature of ES~Cet and the way in which its period is evolving, it should certainly be put to the test. 
 The structure of this paper is the following: Section 2 contains details about observations. In Section 3 are described the average spectra. Section 4 presents information about the trailed spectra. Doppler maps are presented in Section 5. Study of polarisation and pulsations are reported in Section 6 and 7, respectively. The summary and conclusions of our campaign are in Section 8.

%
%

\section{Observations}

We acquired spectroscopy of ES~Cet during two campaigns in 2002 and 2003. During the first run, 528 spectra were obtained on two nights from 2002 October 27 to 28.
The Boller \& Chivens Spectrograph (B\&C) was used to acquire low-resolution spectra on the 6.5-m Magellan-Clay telescope at Las Campanas Observatory, and covered the range from $3850$ to $5500$ \AA. A 0.7-arcsec slit and a 1200 line mm$^{-1}$ grating yielded a spectral resolution of 2 pixels and a dispersion of 0.80 \AA\, pixel$^{-1}$. 
A wide-slit exposure of the spectrophotometric standard star LTT377 was used to obtain a nominal count to flux calibration for the spectra.

The second run was conducted on 2003 October 28 using the FORS1 spectrograph on the VLT telescope. This time 230 spectra were obtained with the range from $4300$ to $8600$ \AA.\, A $1.0$-arcsec slit and a grism GRIS300V yielded a spectral resolution of 2 pixels and a dispersion of 5.30 \AA\,pixel$^{-1}$ for 2x2 binning. 

Table \ref{tab:log_observations} gives an overview of our observations. All images were bias-subtracted  and  flat  field-corrected. Extraction and calibration were carried out using the software packages PAMELA and MOLLY\footnote{http://deneb.astro.warwick.ac.uk/phsaap/software/} \citep{Marsh1989}.

\begin{table}
    \centering
    \caption{Log of ES~Cet observations.}
    \begin{tabular}{l l c c l}
    \hline
    \noalign{\smallskip}
    Telescope & Date & Exp & UT interval &  No. of  \\
              &      & (sec) &             &  frames  \\
    \noalign{\smallskip}
    \hline
    \noalign{\smallskip}
    Magellan & 27 Oct 2002 & 30 - 40 & 03:04 - 08:04 & 357\\
             & 28 Oct 2002 & 30 - 60 & 04:37 - 08:15 & 171\\
    \noalign{\smallskip}
    VLT   & 28 Oct 2003  & 60 & 00:03 - 08:23 & 230\\ 
    \noalign{\smallskip}
    \hline
    \end{tabular}
    \label{tab:log_observations}
\end{table}


\section{Average spectra}
\label{sec:average}

\begin{figure*}
\hspace*{\fill}
\includegraphics[width=0.9\columnwidth]{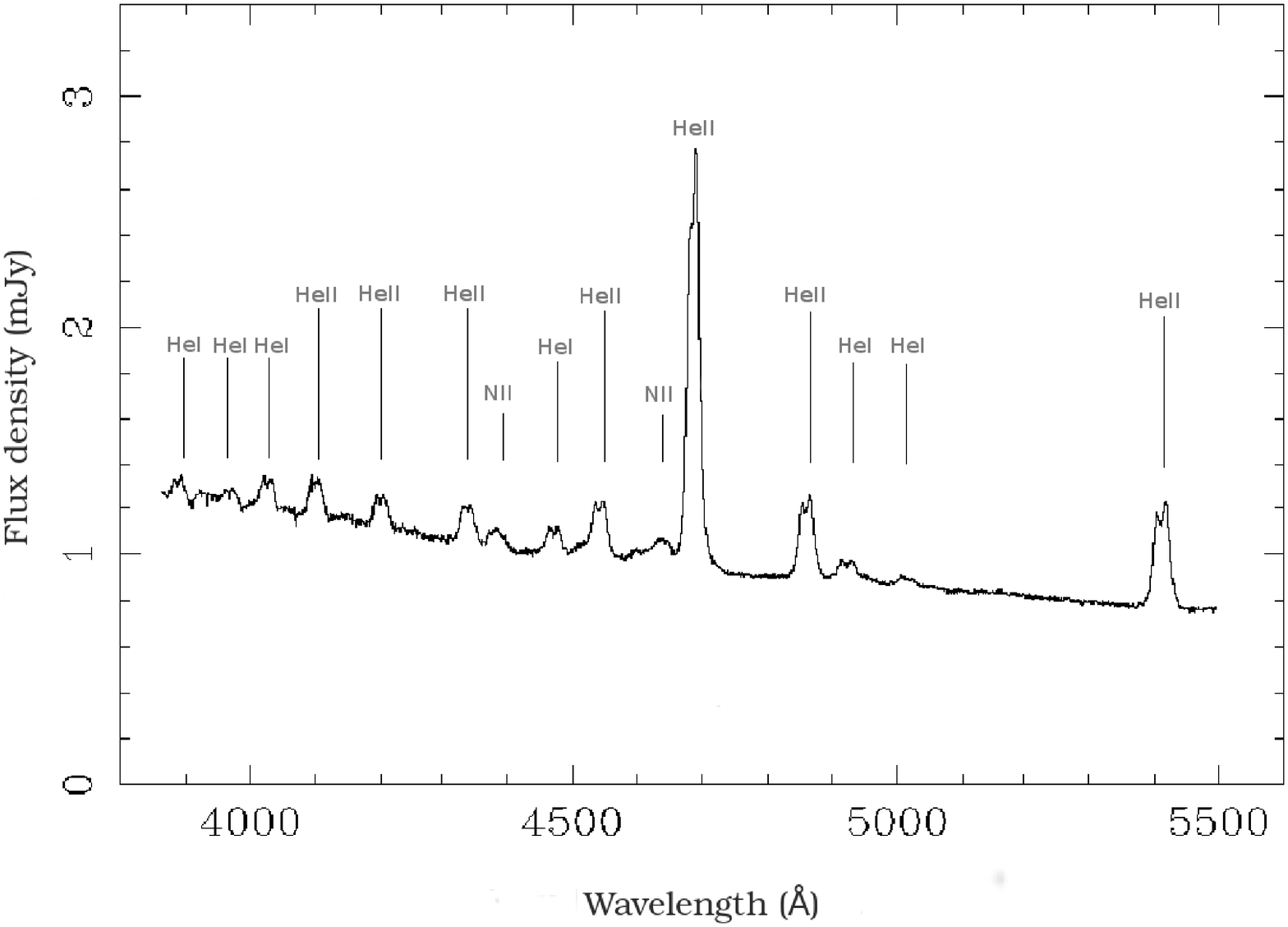}
\hspace*{\fill}
\includegraphics[width=0.9\columnwidth]{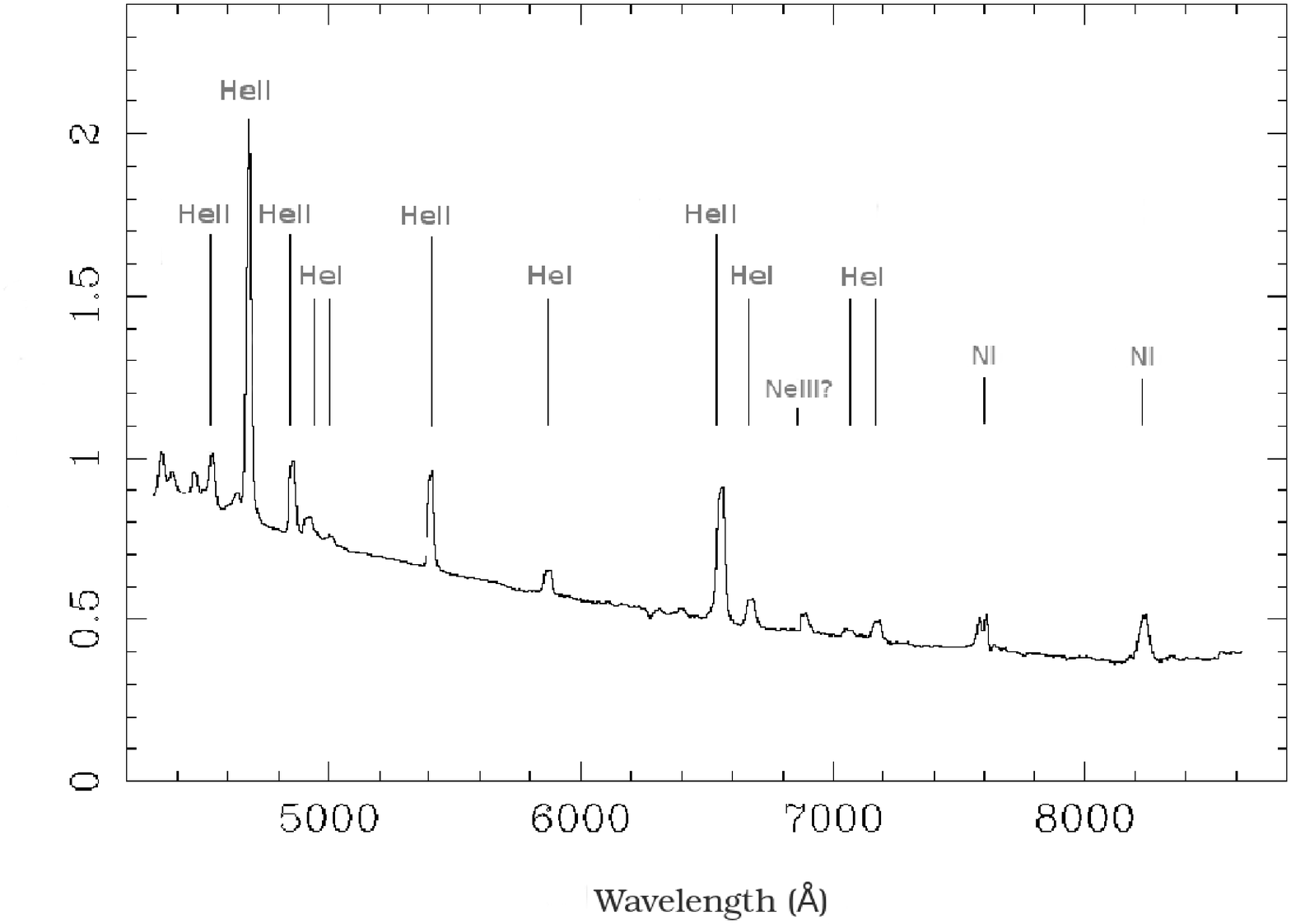}
\hspace*{\fill}
\caption{The panels show the mean spectra of ES~Cet taken with Magellan
27-28 Oct 2002 (left panel) and the VLT on 28 Oct 2003 (right panel). All lines can be
identified with HeII, HeI and NI or NII.}
\label{fig:mean}
\end{figure*}

The average Magellan and VLT spectra of ES~Cet are shown on the left and right panels in Fig.~\ref{fig:mean}, respectively. They consist of a blue continuum with a series of strong emission lines. Most of the identifiable features can be interpreted as either neutral (HeI) or ionized (HeII) helium. The lines from neutral helium are weaker than the ionized helium peaks. Neither hydrogen nor metallic lines are detectable, apart from nitrogen. Nitrogen-rich matter is produced in the CNO-cycle, hence nitrogen seems to be abundant in most AM~CVn stars (e.g.\ \citealt{Marsh1991, Roelofs2009,  Carter2014a, Kupfer2016}). 

\begin{figure}
\hspace*{\fill}
\includegraphics[width=0.85\columnwidth]{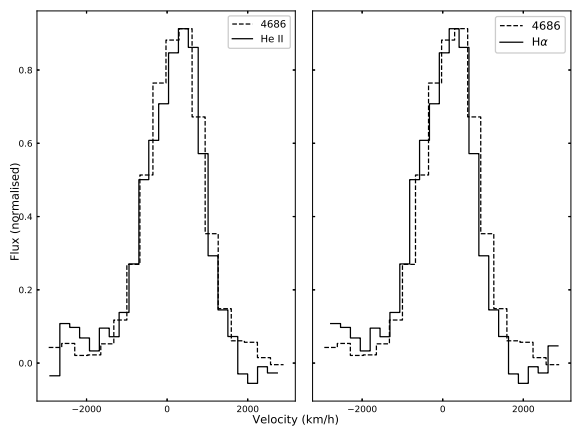}
\hspace*{\fill}
\caption{The emission lines of ES~Cet at 4686\AA~(dotted line) and 6560\AA~(black line). The 6560\AA~line was converted into velocity space treated as HeII line with a central wavelength of 6560.10\AA~(left panel), and as H$\alpha$ line with a central wave length of 6562.72\AA~(right panel). Both lines are shifted but the discrepancy is more significant for H$\alpha$ case (see Sec.\ref{sec:average}).}
\label{fig:gaussian}
\end{figure}

To test for the presence of hydrogen in ES~Cet, we plotted the 6560~\AA\ line as a function of velocity two times over, first relative to the central wavelength corresponding to the HeII line, and second for H$\alpha$ line. We compare these to the velocity profile of HeII~4686 in Fig.~\ref{fig:gaussian}. It is evident that the profiles are better aligned on the assumption that the 6560 line also comes from ionised helium rather than hydrogen. There is thus no evidence for hydrogen in ES~Cet. We also measured the radial velocities of ES~Cet from the HeII~4685.75 line, and from the wavelength corresponding to the 6560.10 HeII line and to the 6562.72 H$\alpha$ line. The mean difference between the $6560.10 - 4685.75$ lines is $-56.1 \pm 1.8$ km/s, while for the $6562.72 - 4685.75$ case this would be $175.9 \pm 1.8$ km/s. Neither is consistent with zero (not surprising because there are some profile differences), but the HeII identification is clearly favoured. 

Our measurements of line fluxes and equivalent widths are given in Table~\ref{tab:lines}. 
The majority of these spectral lines have a broad, double-peaked profile, clearly seen in spectra of ES~Cet, e.g. in the HeII 5411 line. This is a key feature of binary systems with accretion. In some AM~CVn stars, triple-peaked emission lines are observed, i.e. GP~Com \citep{Marsh1999}, SDSS J120841.96+355025.2, SDSS J152509.57+360054.50 and SDSS J012940.05+384210.4  \citep{Kupfer2013}, SDSS J113732.32+405458.3, and SDSS J150551.58+065948.7 \citep{Carter2014}.  \cite{Smak1975} and \cite{Nather1981} suggested that the lines are made up of a double-peaked profiles from an accretion disc, and a separate narrow component near the centre of the line known as a 'central spike'. This feature is thought to originate close to the surface of the accreting white dwarf \citep{Marsh1999,Morales2003}. There is no evidence for the 'central spike' feature in any of ES~Cet's emission lines.  

\begin{table*}
\centering
\caption{Mean fluxes and equivalent widths of lines of ES~Cet.}
\begin{threeparttable}
\begin{tabular}{lcclcc}
\hline   
\noalign{\smallskip}
\multicolumn{3}{c}{Magellan} & \multicolumn{3}{c}{VLT} \\
Line      &     Flux         & EW     & 
Line   & Flux      & EW \\
          & $10^{-15}\,\mathrm{ergs}\,\mathrm{s}^{-1}$ & \AA &
          & $10^{-15}\,\mathrm{ergs}\,\mathrm{s}^{-1}$ & \AA \\
\noalign{\smallskip}
\hline
\noalign{\smallskip}
HeI  3888 & $2.6 \pm 0.1$  & $1.12 \pm 0.05$      & HeII 4541 & $9.5\pm 0.2$   & $6.50 \pm  0.10$   \\   
HeI  3964 & $3.8 \pm 0.1$  & $1.53 \pm 0.05$      & HeII 4686 & $42.2 \pm 0.2$ & $38.2 \pm  0.1$    \\  
HeI  4026 & $5.7 \pm 0.1$  & $2.55 \pm  0.05$     & HeII 4860 & $8.5 \pm 0.1$  & $8.74 \pm  0.05$   \\          
HeII 4100 & $6.7 \pm 0.1$  & $3.20 \pm  0.05$     & HeI  4921 & $3.1 \pm 0.1$  & $3.32 \pm  0.05$   \\  
HeII 4200 & $5.1 \pm 0.1$  & $2.66 \pm  0.05$     & HeI 5015 & $0.89 \pm 0.1$ & $1.01 \pm  0.05$   \\      
HeII 4339 & $6.6 \pm 0.1$  & $3.99 \pm  0.05$     & HeII 5411 & $9.68 \pm 0.1$ & $14.4 \pm  0.05$   \\ 
NII 4379 & $3.6 \pm 0.1$  & $2.25 \pm  0.05$      & HeI  5876 & $2.2 \pm 0.1$  & $4.38 \pm  0.05$   \\
HeI  4471 & $5.0 \pm 0.1$  & $3.37 \pm  0.05$      & HeII 6560 & $50.1\pm 0.2$  & $31.4 \pm  0.1$    \\ 
HeII 4541 & $12.1\pm 0.2$  & $8.51 \pm  0.1$       & HeI 6678 & $2.3 \pm 0.05$  & $7.04 \pm  0.1$  
\\
NII 4643 & $7.0 \pm 0.2$  & $5.23 \pm  0.1$        & NeIII 6886? & $1.26 \pm 0.05$ & $4.32 \pm  0.1$  \\
HeII 4686 & $55.1 \pm 0.2$ & $43.5 \pm  0.1$        & HeI 7065  & $7.2 \pm 0.2$  & $2.7 \pm 0.1$      \\
HeII 4860 & $11.7 \pm 0.1$ & $10.23 \pm 0.05$       & HeI 7160 & $1.7 \pm 0.2$  & $6.89 \pm 0.2$    \\
HeI  4921 & $3.4 \pm 0.1$  & $3.05 \pm  0.05$       & NI 7608\tnote{a} & - & -  \\
HeI  5015 & $1.3 \pm 0.1$  & $1.25 \pm  0.05$       & NI 8242 &$3.4 \pm 0.2$ & $20.74 \pm 0.2$ \\
HeII 5411 & $13.0 \pm 0.1$ & $16.6 \pm  0.1$        & & & \\     
\noalign{\smallskip}
\hline     
\noalign{\smallskip}
\end{tabular}
   \begin{tablenotes}
     \item[a] Line present but could not be measured reliably, due to a blend with another line.
   \end{tablenotes}
\end{threeparttable}
\label{tab:lines}
\end{table*}


\section{Trailed spectra}

\begin{figure}
\hspace*{\fill}
\includegraphics[width=0.9\columnwidth]{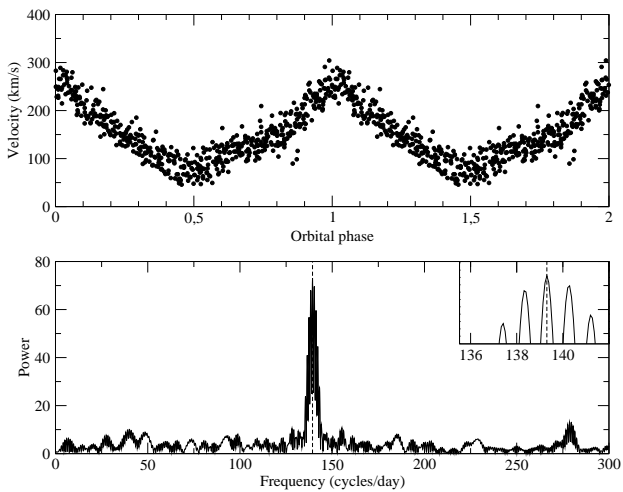}
\hspace*{\fill}
\caption{Top panel shows the radial velocities of the disc through the "S"-wave measured from HeII 4686 line folded on the best fitting period. The bottom panel provides the Lomb-Scargle periodogram for the same line with a vertical dashed line showing the position of the Copperwheat's et al's ephemeris. A zoom presents a magnified view of the strongest peak.}
\label{fig:RV}
\end{figure}

We present the phased-binned trailed spectra for the strongest HeII and HeI lines of ES~Cet on the upper panels in Fig.~\ref{fig:maps1} and Fig.~\ref{fig:maps2}. Twenty phase bins were equally spaced around the cycle and we have displayed one cycle repeated twice for clarity. The time-resolved spectra were folded on the ephemeris given by \cite{Copperwheat2011} (this spans the epochs of our spectra):
\[ \textnormal{BMJD(TDB)}=52200.980575(6) + 0.007178375 98(3)\,E, \]
where $E$ is the cycle number. 
The trailed spectra show clear evidence for double-peaked emission lines with a sinusoidal "S"-wave component superposed in most instances.  Therefore, these data confirm that the 620\,s photometric period is indeed orbital, beyond any doubt. To confirm this quantitatively, we measured radial velocities and computed the Lomb-Scargle periodogram corresponding to the strong HeII~4686 line (Fig.~\ref{fig:RV}). A clear signal is seen at $~139$ cycles per day. A weaker group of peaks appears at 2 times this frequency. Zooming in on the main signal, one can see that the strongest peak occurs at $f=139.306725$ cycles per day ($P_{orb}=0.007178404$ days), with the two next-strongest peaks occurring at the usual $\pm1$ cycle per day aliases. The period of the strongest peak corresponds to the one presented by \cite{Copperwheat2011}, and hence represents the orbital period of the binary. 

\begin{figure}
\hspace*{\fill}
\includegraphics[width=0.9\columnwidth]{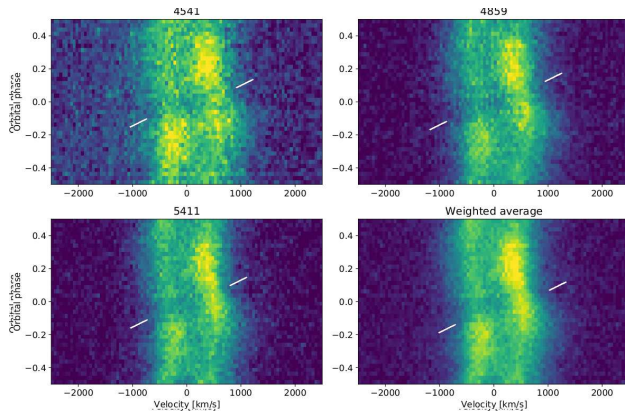}
\hspace*{\fill}
\caption{The closeup of the profiles near eclipse in trailed spectra of three of the lines from the Magellan data: HeII 4541 (top left), HeII 4859 (top right), HeII 5411 (bottom left) and the average of these three lines (bottom right). The eclipse is indicated by the white lines and is visible as a darkening around phase~0.0 that slopes upwards from the blue- to the red-shifted sides of the lines.}
\label{fig:eclipse}
\end{figure}

In the upper panels of Figs~\ref{fig:maps1} and \ref{fig:maps2}, one can see a dark, near-horizontal line on the blue-shifted side of the HeII~4859, HeII~5411, HeII~4339 and HeII~4541 trailed spectra. HeII~4859 shows it best of all, and shows that it extends to the red-shifted side of the line, although bright-spot emission dominates on the red-shifted side beyond $+400$\,km/s. The feature, which is shown magnified in Fig.~\ref{fig:eclipse} (indicated by the white lines), has all the signs of a "rotational disturbance" \citep{Greenstein1959} caused as the disc is eclipsed by the mass donor. For instance, it occurs earliest in phase on the blue-shifted peak of the emission line, as expected for a prograde orbiting disc. The feature is seen close to phase zero on \cite{Copperwheat2011}'s ephemeris which is referenced to minimum light. This indicates that the photometric signal is primarily caused by eclipses of the disc. This association is further confirmed by the phasing of the S-wave which is in a standard location relative to phase zero in the trailed spectra, as also confirmed by the location of the bright emission in the Doppler maps. 

Thus, we conclude that ES~Cet's outer disc is eclipsed. The eclipse can be traced as far as $\sim -800$\,km/s (Fig.~\ref{fig:eclipse}). Given that the outermost velocity in the disc is around $600$\,km/s from the emission line peak velocity, and that $V \propto R^{-1/2}$ in a Keplerian disc, the eclipse therefore reaches a radius in the disc which is a fraction $(600/800)^2 \approx 50$\% of the outer disc radius. The signal-to-noise ratio in the line wings prevents us from determining whether the eclipse reaches the white dwarf itself, although this seems unlikely given the 10 to 20\% depth of the photometric eclipses \citep{Copperwheat2011}.
Our spectra had exposure times of 60\,s, corresponding to $\sim 0.1$ in terms of orbital phase, very comparable to the vertical extent of the eclipse feature in Fig.~\ref{fig:eclipse} at any one velocity in the lines. The data are therefore in effect vertically smeared by this amount, and so shorter exposure spectra have the potential to reveal deeper and sharper variations which may allow the orbital inclination of ES~Cet to be pinned down. The eclipse may also encompass the bright spot, as is perhaps best indicated by the HeII 4859 trail in Fig.~\ref{fig:eclipse}.

The various trails are similar, but not identical, and HeII~4686 in particular is noticeably asymmetric from red- to blue-shifted sides. The S-wave from the bright-spot also seems most visible at its extremes of radial velocity. These features may hint at vertical structure and self-obscuration within the disc. This may not be surprising as ES~Cet is viewed at high inclination, and it is also a high accretion rate system.


\section{Doppler maps}

To accurately track the phase of the S-wave signal, we back-projected the trailed spectra into a Doppler tomogram \citep{Marsh1988}. The method of Doppler tomography enables the projection of a series of phase-resolved spectra on to a two-dimensional map in velocity coordinates (for review: \citealt{Steeghs2003, Marsh2001}).  In Doppler tomograms, emission features that are not stationary in the binary frame or move on a period different from the orbital period will spread out over the resulting Doppler tomogram, while stationary emission features add up constructively. Not only is the method of Doppler tomography useful to separate features that move with a different amplitudes and/or phase, but it allows tracking of asymmetric structures in accretion discs and reveals details of the gas flow in a variety of systems. For analysis of AM CVn systems, Doppler tomography has also proved to be beneficial,  e.g.\ for the investigation of the structure of an accretion disc \citep{Breedt2012}, the discovery of multiple bright spots \citep{Roelofs2006a,Kupfer2013}, or for searching for the presence of central spikes \citep{Kupfer2016}. 

\begin{figure*}
\hspace*{\fill}
\includegraphics[width=0.88\textwidth]{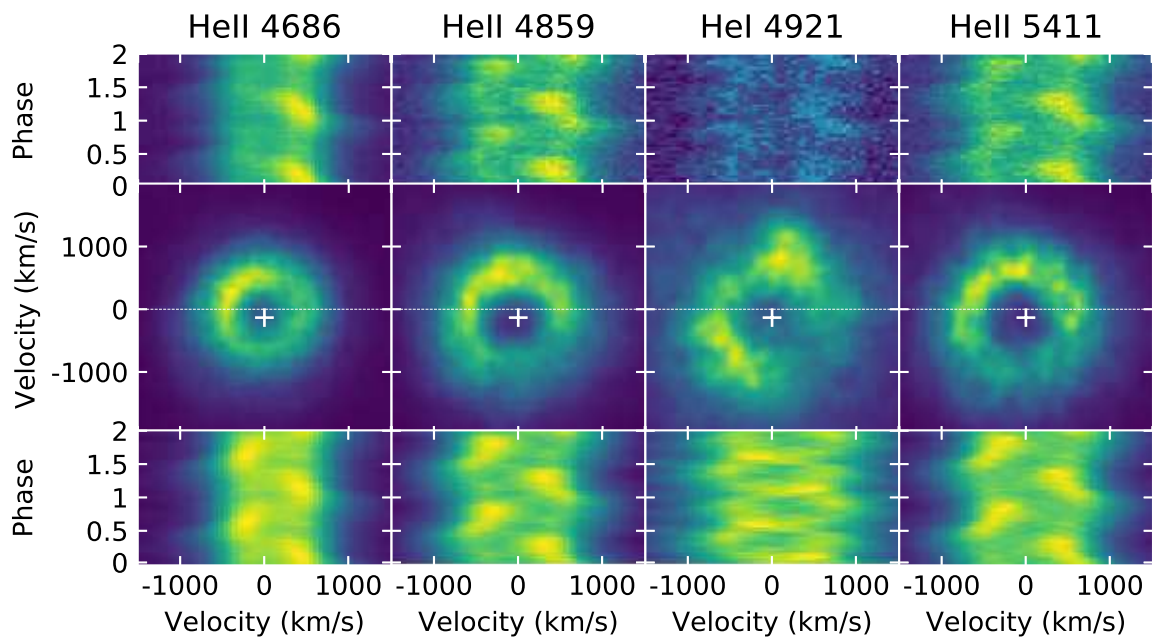}
\hspace*{\fill}
\vspace{-1cm}
\caption{The top panels show phase-folded, continuum-normalised and subtracted trailed spectra of four of the lines from the Magellan data. The middle and bottom panels show the equivalent Doppler maps and fits. White crosses mark the center of a disc. At phase~1, the HeII~4859 data show a dark feature, especially on the blue-shifted side, and slightly delayed in phase near zero velocity, that is the signature of an eclipse of the disc.}
\label{fig:maps1}
\end{figure*}

\begin{figure*}
\hspace*{\fill}
\includegraphics[width=0.88\textwidth]{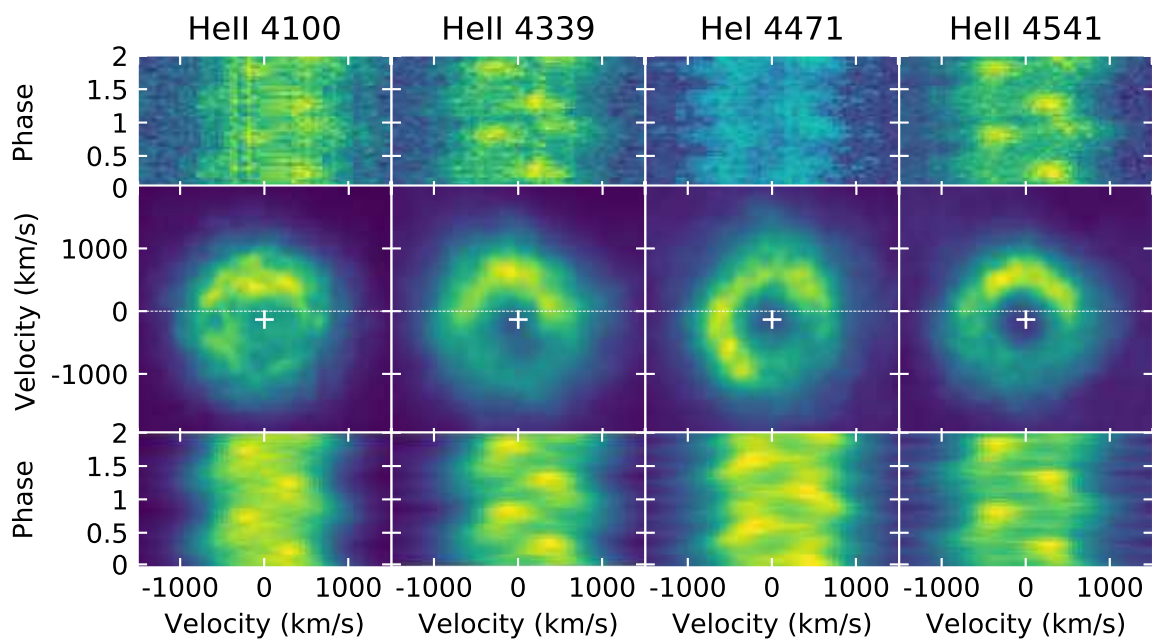}
\hspace*{\fill}
\vspace{-1cm}
\caption{As in Fig.~\protect\ref{fig:maps1}, the trailed spectra and the equivalent Doppler maps and fits for four more lines from the the Magellan data.}
\label{fig:maps2}
\end{figure*}

For all the trailed spectra, we computed the corresponding Doppler tomograms using the software package DOPPLER\footnote{https://github.com/trmrsh/trm-doppler} \citep{Marsh1988}. The tomograms are presented in the second rows of Fig.~\ref{fig:maps1} and Fig.~\ref{fig:maps2}. The bottom panels show trailed spectra computed from the maps. Once again, for each map and trailed spectrum, we used the ephemeris given by \cite{Copperwheat2011}. All the Doppler tomograms show a similar structure, with a bright ring corresponding to emission from the accretion disc. In the case of the bright spot, the situation is more complicated. It is located mostly towards the upper-left, which, as remarked above, is the expected location, but it  has a smeared structure and is not consistent from line to line. Partly this might be azimuthal smearing due to ES~Cet's short period and the 60\,s-long exposures as discussed earlier for the eclipses, but in addition, ES~Cet clearly is a very high accretion rate system, and the Doppler maps of high accretion rate CVs (nova-likes) have often proved hard to unravel, so we do not attach much significance to this finding. Indeed, compared to many nova-likes, which exhibit single-peaked lines even when deeply eclipsing, ES~Cet is well-behaved.

\section{Polarimetry}

Given the strong HeII emission in the magnetic polar (AM~Her) class of CVs, we decided to obtain spectropolarimetry of ES~Cet. We took one night of circular spectropolarimetry using the FORS1 spectrograph of the VLT on 28 October 2003. We took spectra for slightly over 8 hours, obtaining 230 exposures of 60 seconds each, the remaining time taken up with overheads of rotating the Wollaston prism. 

\begin{figure}
\hspace*{\fill}
\includegraphics[width=0.9\columnwidth]{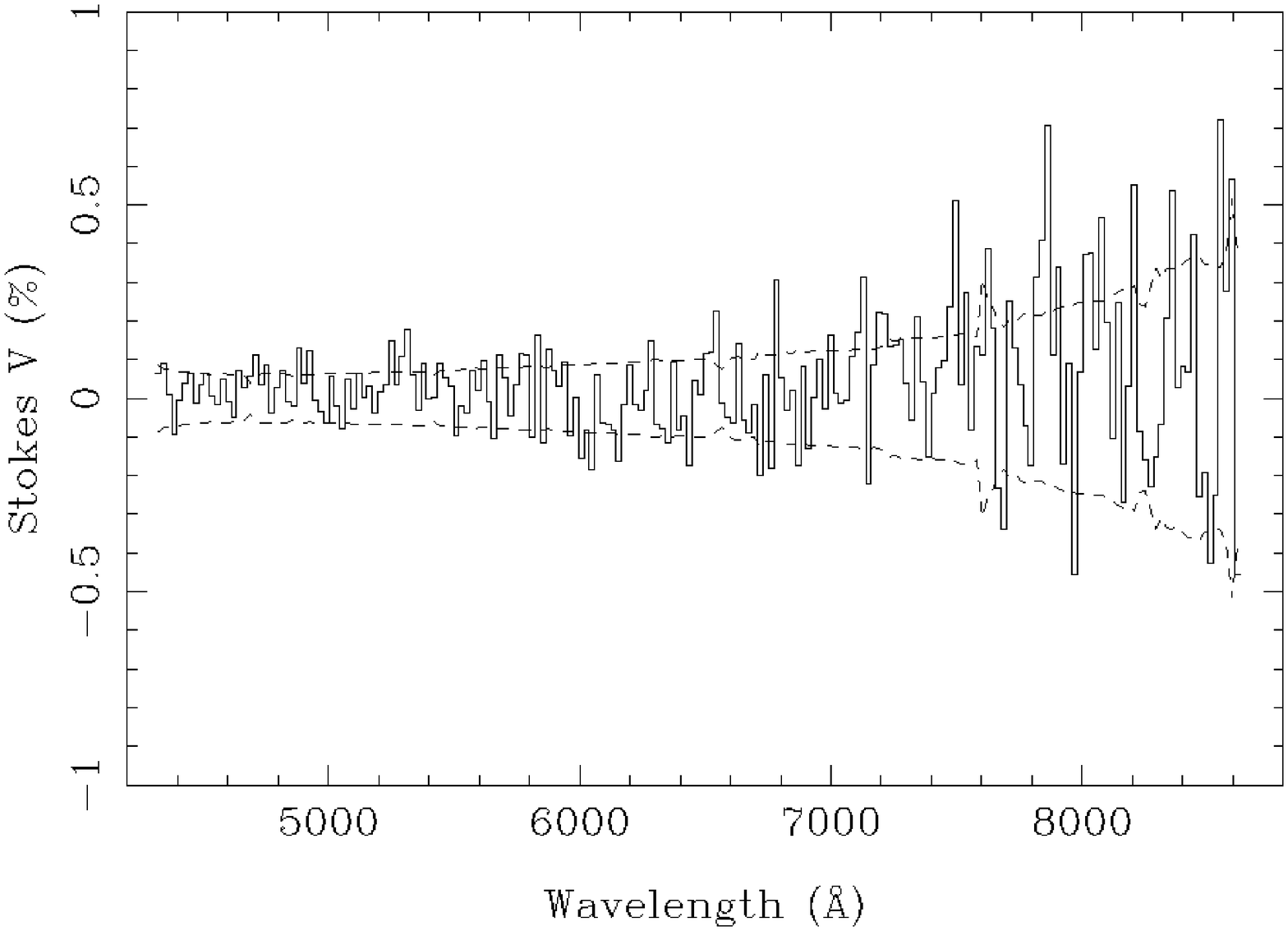}
\hspace*{\fill}
\caption{The mean percentage circular polarisation of ES~Cet observed
  with the VLT on 28 October 2003. The dashed
  lines show the estimated uncertainties ($1\sigma$).}
\label{fig:pol_spec}
\end{figure}

\begin{figure}
\hspace*{\fill}
\includegraphics[width=0.9\columnwidth]{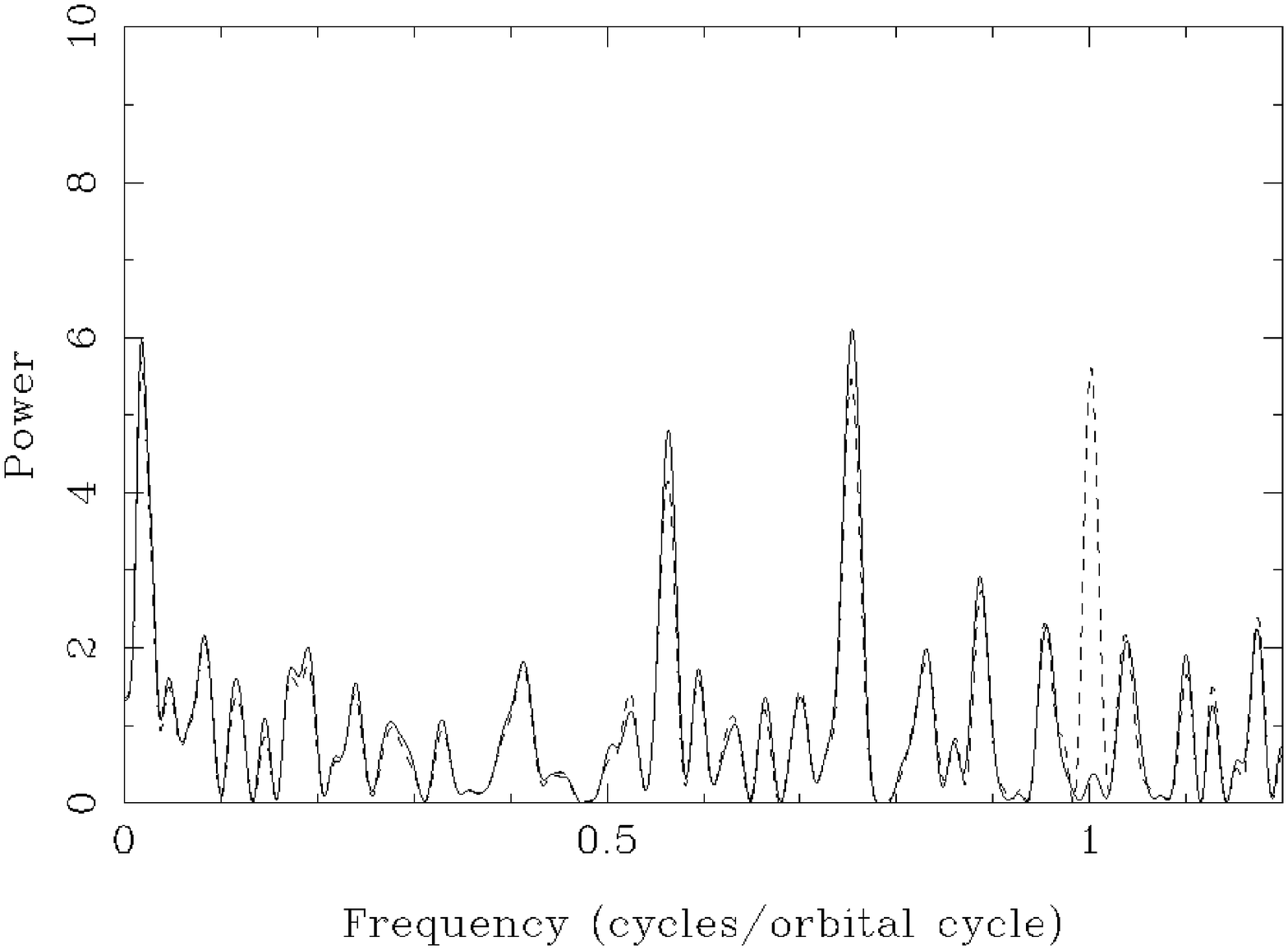}
\hspace*{\fill}
\caption{The Lomb-Scargle periodogram of the light curve of the circular polarisation integrated from 4400 to 7000 \AA. Any signal on the orbital period should appear at $1\,$cycle/orbital cycle. The dashed line shows the periodogram after the addition of a signal with a semi-amplitude polarisation of $\sim 0.05$\,\% at the expected frequency.}
\label{fig:pol_scargle}
\end{figure}

The average polarisation is consistent with zero, with an uncertainty of order $0.1$\,\% (Fig.~\ref{fig:pol_spec}).
To check that this could not be the result of cancellation of opposing senses of circular polarisation, we also calculated the power spectrum of the polarisation light curve integrated from $4400$ to $7000$ \AA\ (Fig.~\ref{fig:pol_scargle}). 
This is also a null result, and in this case we would have expected to detect a semi-amplitude of $0.05$\,\%. We illustrate this by injection of such a signal at the expected frequency (dashed line in Fig.~\ref{fig:pol_scargle}) which produces a peak of height $\sim 6$, which has a probability (frequency known in advance) of $\exp(-6) \sim 0.25$\,\%. We conclude that there is no circular polarisation in ES~Cet at a level of $0.05$ to $0.1$\,\%, and thus no direct sign of magnetic accretion; the mystery of the absence of signs of magnetism amongst AM~CVn stars continues.


\section{Discussion}

Several models have been proposed to explain the
observed properties of the two shortest period AM~CVn stars, HM~Cnc and V407~Vul \citep{Ramsay2000,Israel2003,Barros2007}. The intermediate polar (IP) model \citep{Motch1996} model holds that these systems are not ultra-compact binaries at all, but rather have orbital periods of several hours and the ultra-short periods then represent the spins of magnetic white dwarfs. The unipolar induction (UI) model \citep{Wu2002} model is the only model without a Roche lobe filling secondary and is essentially equivalent to the model proposed by \cite{Goldreich1969} for the Jupiter–Io system. In the direct-impact accretion model \citep{Marsh2002} a Roche lobe-filling white dwarf loses its mass to its more massive white dwarf companion, and the accretion stream hits the accretor directly without forming a disc. This is presently the most widely-accepted model of HM~Cnc and V407~Vul \citep{Steeghs2006,Wood2009,Roelofs2010}. The natures of AM~CVn stars with longer orbital periods, e.g. SDSS J135154.46-064309.0 ($P_{orb}=15.7$ min, \citealt{Green2018}) and AM~CVn itself ($P_{orb}=17.1$\,min, \citealt{Nelemans2001c}), are better established. Their spectra show an absence of hydrogen, the presence of helium lines, many of which are the double-peaked emission (or absorption) lines characteristic of sources accreting via discs. Also, detection of one or even two bright spots are frequently reported \citep{Kupfer2016}. Based on the photometric observations, \cite{Espaillat2005} concluded that ES~Cet follows the direct-impact scenario and is therefore similar to HM~Cnc and V407~Vul. However, we find instead clear evidence of an accretion disc along with a smeared spot-like structure, placing it very much in the same bracket as the bulk of longer period AM~CVn systems.

Although, ES Cet has a disc and is more like the other ultra-compact binaries, AM~CVn stars at longer periods than ES~Cet tend to have very different spectra with weak absorption lines, e.g. HP Lib \citep{Roelofs2007}. AM~CVn itself \citep{Roelofs2006b} shows an absorption spectrum dominated by HeI, with only HeII~4686 in emission whereas ES~Cet is very much an emission line system without any presence of absorption lines. This suggests a difference in the vertical temperature structure within the disk. That is, in AM~CVn one sees light from a region where the temperature drops along the line of sight towards the observer, whereas in ES~Cet it increases. The high orbital inclination implied by the eclipses may also play a role. Support for this comes from \cite{Burdge2020}'s discovery of ZFT J1905+3134, an eclipsing AM~CVn system with a period of $\sim 17.2$ minutes which also shows strong HeII emission features similar to ES~Cet. 


\section{Conclusions}

We have described results of Magellan and VLT spectroscopic and spectropolarimetric surveys aimed at understanding the internal accretion structure of the helium-rich binary ES~Cet. 

Time-resolved spectra reveal strong variability in the emission lines on the 620\,s period found from photometry. Double-peaked emission is clearly visible in the lines in our spectra, as is characteristic of an accretion disc. This confirms that the photometric period of 620\,s first reported by \cite{Warner2002} is ES~Cet's orbital period. We find no periodic signals on periods unrelated to the 620\,s orbital period. 

Due to the strong HeII emission in magnetic CVs, which is also seen in ES~Cet, we conducted one night of circular spectropolarimetry, but we found no circular polarisation placing an upper limit of $0.1\,\%$. The strength of HeII in ES~Cet more likely reflects its high accretion rate and state of excitation.

We discovered a short-lived, phase-dependent flux deficit in the trailed spectra of the HeII~4859, HeII~5411, HeII~4339 and HeII~4541 lines, consistent with a "rotational disturbance" \citep{Greenstein1959}. This shows that the outermost parts of the disc in ES~Cet are eclipsed by the mass donor. The timing of the deficit is consistent with the phase of photometric minimum light, indicating that a significant part of the orbital light curve may be caused by eclipses. The eclipse extends over about half the outer disc in radius, but probably does not reach the white dwarf itself given the relatively shallow photometric eclipse. Higher time resolution spectroscopy of ES~Cet could refine the exact radial extent of the eclipse and therefore constrain the orbital inclination.

For all the trailed spectra, we computed the corresponding Doppler tomograms. All the Doppler tomograms show evidence for a similar structure, with a bright ring corresponding to emission from the  accretion disk. Several also show a brightness maximum at the expected location of the gas stream / disc impact ("bright spot"). 

We conclude that, with the third shortest orbital period of any AM~CVn star ($P_{orb}=620$\,s), ES~Cet is the system with the shortest orbital period that hosts an accretion disc, and that the disc is eclipsed.


\begin{acknowledgements}
      Project was supported by Polish National Science Center grants awarded by decisions: DEC-2015/16/T/ST9/00174 for KB. TRM and DS acknowledge support from the Science and Technology Facilities Council (STFC) grant numbers ST/P000495/1 and ST/T000406/1. Based on observations collected at the European Organisation for Astronomical Research in the Southern Hemisphere under ESO programme 072.D-0119(A) as well as data gathered with the 6.5 meter Magellan Telescopes located at Las Campanas Observatory, Chile.
\end{acknowledgements}


\end{document}